# Simple ROI untuk justifikasi investasi proyek Data Warehouse pada perguruan tinggi swasta

**Spits Warnars**

Department of Computing and Mathematics ,Manchester Metropolitan University
John Dalton Building, Chester Street, Manchester M1 5GD, United Kingdom
s.warnars@mmu.ac.uk

**Abstract.** Decreasing new students for private high education push the management particularly for high level management for making an information which can help them to make decisions in order for competition with other high educations. One of way out by building with information technology approaching like data warehouse for data handling and making the best decisions. Simple ROI is used for project justification. Based on ROI value between 1,850.13% and cash flow Rp. 22,081,297,308 then can be concluded that project data warehouse development in private high education can be implemented with the particular assumptions.

**Abstrak.** Berkurangnya jumlah mahasiswa baru untuk perguruan tinggi swasta memaksa manajemen khususnya manajemen tingkat atas untuk berpaling guna membuat sebuah informasi yang dapat membantu mereka dalam membantu mengambil keputusan dalam rangka berkompetisi dengan perguruan tinggi lainnya. Salah satu jalan keluarnya adalah dengan membangun dengan pendekatan teknologi informasi seperti data warehouse untuk mengelola data dan memberikan pembuatan pengambilan keputusan yang paling terbaik. Simple ROI digunakan untuk menilai kelayakan proyek. Berdasarkan nilai ROI yang berkisar 1850.13% dan nilai total aliran uang kas yang mencapai Rp. 22,081,297,308, dapat disimpulkan bahwa proyek pengembangan data warehouse pada perguruan tinggi swasta layak untuk diimplementasikan dengan asumsi-asumsi yang ada.

## 1 Pendahuluan

Meningkatnya jumlah perguruan tinggi swasta tak pelak menimbulkan persaingan antara perguruan tinggi swasta tersebut dan tantangan yang terberat adalah dengan perguruan tinggi negeri yang terlebih dulu telah ada dan didukung oleh pemerintah. Sasaran persaingan yang mengedepankan sistem informasi sebagai sebuah alat teknologi informasi untuk memenangkan persaingan adalah salah satu alternatif yang dapat dipilih di jaman informasi ini dan di tangan yang tepat informasi akan menjadi sebuah senjata ampuh untuk dapat memenangkan persaingan ini dalam mendapatkan mahasiswa baru sebagai salah satu sumber pendapatan untuk perguruan tinggi swasta dan minat dari masyarakat.

Manajemen tingkat atas sebagai bagian dalam pengambilan keputusan strategis kerap kali tidak didukung dengan informasi pelaporan yang seharusnya dan terkesan seadanya dan dibuat-buat, padahal apa yang manejemen tingkat atas tersebut putuskan adalah sebuah keputusan strategis yang mempengaruhi proses bisnis pada organisasi bisnis tersebut. Manajemen tingkat atas pada perguruan tinggi seringkali diabaikan dalam mendapatkan haknya didukung oleh sistem informasi pengambilan keputusan yang membantu tugas-tugas mereka dalam mengambil keputusan secara lebih tepat dan akurat.

Datawarehouse adalah bukan sebuah hal yang baru dalam dunia pengambilan keputusan, datawarehouse bukanlah sebuah software ataupun hardware, melainkan hanyalah sebuah sistem informasi yang memanfaatkan teknologi informasi untuk mendukung pengambilan keputusan guna kepetingan manajemen khususnya manajemen tingkat atas. Datawarehouse hanyalah salah satu alternatif dari sekian banyak sistem pengambilan keputusan dan pemanfaatan pemakaian datawarehouse haruslah didukung pula oleh manajemen tingkat atas dalam penerapannya. Tanpa dukungan manajemen tingkat atas pemanfaatn datawarehouse hanyalah sebatas cerita-cerita di buku saja dan tidak akan berguna di kemudian hari.

Seringkali proyek yang berhubungan dengan teknologi informasi selalu dihubungkan dengan hal-hal yang memboroskan uang dan tidak kelihatan nilai investasinya, akibatnya banyak manajemen tingkat atas yang kurang berpengalaman sering melihat ini sebagai bagian dari menolak untuk penerapan proyek teknologi informasi. Oleh karena itu untuk menerapkan sebuah proyek teknologi informasi kiranya perlu dilakukan sebuah studi dan analisa yang hasilnya dapat memberikan penilaian dan gambaran bagi manajemen tingkat atas.

Analisa dampak investasi Teknologi Informasi digunakan untuk menilai apakah pembuatan proyek Data Warehouse pada perguruan tinggi layak untuk diimplementasikan. Nilai arus kas bersih didapatkan dengan mengurangkan jumlah manfaat bersih dengan jumlah biaya, yang pada akhirnya persentase ROI (Return On Investment) akan didapatkan dengan cara membagi dengan jumlah tahun dan dibagi lagi dengan nilai investasi proyek [11] [8].

## 2 Analisa Manfaat

Untuk mendapatkan prosentase ROI digunakan rumus berikut:

ROI = total manfaat - total biaya / 5 tahun / Nilai bersih investasi proyek

Berikut ini merupakan manfaat-manfaat laporan sebagai hasil keluaran dari sistem yang berjalan dan yang digunakan oleh manajemen perguruan tinggi yaitu:

1) Mengurangi biaya administrasi pembuatan laporan

2) Mengurangi tenaga pembuatan laporan

3) Mempercepat waktu pembuatan laporan/administrasi

4) Meningkatkan jumlah mahasiswa baru

5) Meningkatkan bantuan dana dari pihak ketiga (ADB, proyek TPSDP Bank Dunia)

6) Meningkatkan produktivitas manajemen tingkat atas

7) Meningkatkan citra perguruan tinggi

8) Meningkatkan hubungan dengan stakeholder (mahasiswa, orang tua, dunia kerja, DIKTI)

9) Meningkatkan moral karyawan

10) Meningkatkan pengetahuan manajemen

11) Meningkatkan perencanaan pengelolaan data

12) Meningkatkan fleksibilitas pemanfaatan data

13) Meningkatkan kemampuan pengambilan keputusan

Ada dua tipe manfaat teknologi informasi atau sistem informasi yang umum dipakai yaitu manfaat yang terlihat (Tangible benefit) dan manfaat yang tidak terlihat (Intangible benefit). Manfaat tersebut dapat diukur (measurable benefit) dan juga sulit diukur (immeasurable benefit)[3]. Berdasarkan kategori tersebut manfaat-manfaat diatas disarikan dalam tabel matrik manfaat berikut :

Tabel 1 Matrik Manfaat

| | | | |
|---|---|---|---|
| | **Tinggi** | 1. Mengurangi biaya administrasi pembuatan laporan<br>2. Mengurangi tenaga pembuatan laporan | 3. Mempercepat waktu pembuatan laporan/administrasi |
| **T**<br>**a**<br>**n** | **Rendah** | 4. Meningkatkan jumlah mahasiswa baru<br>5. Meningkatkan bantuan dana dari pihak ketiga (ADB, proyek TPSDP, Bank dunia)<br>6. Meningkatkan produktivitas | 7. Meningkatkan citra perguruan tinggi<br>8. Meningkatkan hubungan dengan stakeholder (mahasiswa, orang tua, dunia kerja, DIKTI)<br>9. Meningkatkan moral karyawan<br>10. Meningkatkan pengetahuan |
| g i b l e | | manajemen tingkat atas | manajemen<br>11. Meningkatkan perencanaan pengelolaan data<br>12. Meningkatkan fleksibilitas pemanfaatan data<br>13. Meningkatkan kemampuan pengambilan keputusan |
| | | **Tinggi** | **Rendah** |
| | | **Measurable** | |

Selain itu ke-13 manfaat-manfaat tersebut dipetakan menjadi tabel 2:

Tabel 2 Potensi Manfaat

| Potensi Manfaat | Aspek Manfaat | Klasifikasi Domain | Value | Metode pengukuran |
|---|---|---|---|---|
| 1 | Tangible Measurable | Teknologi | Finansial | Simple ROI |
| 2 | Tangible Measurable | Teknologi | Finansial | Simple ROI |
| 3 | Tangible Immeasurable | Teknologi | Non Finansial | - |
| 4 | Intangible Measurable | Bisnis | Finansial | Simple ROI |
| 5 | Intangible Measurable | Bisnis | Finansial | Simple ROI |
| 6 | Intangible Measurable | Bisnis | Finansial | Simple ROI |
| 7 | Intangible Immeasurable | Bisnis | Non Finansial | - |
| 8 | Intangible Immeasurable | Bisnis | Non Finansial | - |
| 9 | Intangible Immeasurable | Bisnis | Non Finansial | - |
| 10 | Intangible Immeasurable | Teknologi | Non Finansial | - |
| 11 | Intangible Immeasurable | Teknologi | Non Finansial | - |
| 12 | Intangible Immeasurable | Teknologi | Non Finansial | - |
| 13 | Intangible Immeasurable | Teknologi | Non Finansial | - |

Terlihat dari tabel 2 hanya manfaat yang dapat diukur (measurable) yang dapat diukur dengan metode analisa manfaat/biaya (simple ROI). Baik manfaat yang dapat diukur (measurable) yang terlihat (Tangible benefit) dan yang tidak terlihat (Intangible benefit). Sedangkan untuk manfaat yang tidak dapat diukur (immeasurable) baik yang terlihat (Tangible benefit) dan yang tidak terlihat (Intangible benefit) tidak dilakukan pengukuran pada proyek pembuatan Data Warehouse pada perguruan tinggi ini [1][4]. Khusus untuk manfaat meningkatkan bantuan dana dari pihak ketiga tidak dilakukan pengukuran yang dikarenakan adanya kesulitan untuk mengukur manfaat yang akan diuraikan, dan masih adanya kurangnya ketegasan manfaat ini dimana manfaat bantuan dana pihak ketiga ini bisa juga didapatkan tanpa membangun proyek pembuatan Data Warehouse ini.

Seluruh perhitungan proyek ini mengasumsikan batasan waktu sampai 5 tahun, yang didasarkan bahwa implementasi sebuah proyek sistem informasi akan dapat bertahan dan dirancang untuk kebutuhan 5 tahun kedepan sesuai dengan keinginan manajemen tingkat atas, sehingga ada kemungkinan 5 tahun kedepan manajemen tingkat atas dapat menilai kelayakan sebuah aplikasi sistem informasi apakah perlu dikembangkan atau dirubah sama sekali.

## 3 Persyaratan Kebutuhan Proyek Data Warehouse

Untuk melihat sejauh mana besarnya proyek pengembangan Data Warehouse ini maka dibawah ini akan diuraikan berapa personal yang terlibat dalam proyek ini, lama dan uraian kegiatan pengerjaan proyek serta biaya yang dibutuhkan untuk mengembangkan proyek ini [9][6]

1) Waktu pengerjaan proyek

Waktu pengerjaan proyek ini memakan waktu selama 1 tahun dan dengan menggunakan tabel 3 yang merupakan gantt chart pengerjaan proyek yang akan diperlihatkan rincian proses kegiatan pengembangan proyek sebagai berikut :

Tabel 3 Gantt Chart Pengerjaan Proyek

| **Proses kegiatan** | **Bulan** | | | | | | | | | | | |
|---|---|---|---|---|---|---|---|---|---|---|---|---|
| | **1** | **2** | **3** | **4** | **5** | **6** | **7** | **8** | **9** | **10** | **11** | **12** |
| Analisa Laporan-laporan atau informasi yang dibutuhkan | ■ | ■ | ■ | | | | | | | | | |
| Analisa model data logika basis data system OLTP | | | ■ | ■ | | | | | | | | |
| Desain Hypercubes | | | | ■ | ■ | | | | | | | |
| Desain model data logika datawarehouse | | | | | ■ | ■ | | | | | | |
| Seminar awal | | | | | | ■ | | | | | | |
| Desain Struktur Data Fisik Datawarehouse | | | | | | ■ | ■ | ■ | | | | |
| Desain Aplikasi ETL Datawarehouse | | | | | | | | ■ | ■ | | | |
| Desain Aplikasi OLAP Datawarehouse | | | | | | | | | ■ | ■ | ■ | |
| Seminar Akhir | | | | | | | | | | | | ■ |

2) Staf proyek yang terlbat

Adapun proyek pengembangan Data Warehouse ini akan melibatkan 6 orang staf proyek dengan rincian 1 orang kepala proyek, 2 orang Data Warehouse administrator, 2 orang programmer dan 1 orang administrasi. Staf proyek akan mempunyai hari kerja ditentukan mulai hari Senin sampai Jum'at terhitung selama 1 tahun yaitu 260 hari kerja sesuai dengan lama proyek ini, dan setiap harinya membutuhkan waktu kerja 7 jam per hari. Dengan demikian jumlah keseluruhan jam kerja yang dibutuhkan untuk setiap orang staf pada proyek ini dalam satu tahun adalah : 7 jam/hari * 260 hari kerja = 1820 jam.

3) Biaya Pengembangan proyek

Proyek pengembangan Data Warehouse ini membutuhkan biaya yang terdiri dari upah staf proyek, pengadaan perangkat lunak, pengadaan perangkat keras, jaringan dan biaya pendukung lainnya. Upah staf proyek dibayar per jam sesuai dengan upah per jam masing-masing staf yang dapat dilihat pada tabel 4.

Tabel 4 Upah Staf proyek

| **Staf** | **Upah/jam** | **Upah/jam * 1820 jan** | **Total bayar** |
|---|---|---|---|
| Kepala Proyek (1) | 20.000 | 36.400.0 | 36.400.000 |
| *Data Warehouse Administrator* (2) | 15.000 | 27.300.0 | 54.600.000 |
| *Programer* (2) | 10.000 | 18.200.0 | 36.400.000 |
| Administrasi (1) | 5.000 | 9.100.0 | 9.100.000 |
| Total keseluruhan upah | | | 136.500.000 |

Untuk biaya pengadaan perangkat lunak, akan terbagi menjadi biaya pembuatan aplikasi dan basis data. Aplikasi akan dibuat dengan menggunakan bahasa pemrograman Java dimana bahasa pemrograman Java ini merupakan bahasa pemrograman open source, jadi biaya untuk pengadaan bahasa pemrograman ini tidak diperlukan Sedangkan basis data akan dibuat dengan menggunakan basis data Oracle, dimana saat ini perguruan tinggi diasumsikan telah beralih ke basis data Oracle 10g sehingga tidak diperlukan biaya pengadaan basis data.

Tabel 5 memperlihatkan biaya pengadaan perangkat keras, table 6 memperlihatkan biaya pengadaan jaringan dan table 7 untuk biaya pendukung

Tabel 5 Biaya pengadaan perangkat keras

| **Hardware** | **Harga** |
|---|---|
| 1 buah server HP Proliant ML110G2-063 Intel Pentium 4 processor 540(3.2 GHz, FSB DD, cache 1MB) Memory 512MB DDR400 ECC, Single Channel U320 SCSI, Hard drive 73 GB SCSI U320 10K., 48x CD, 8MB VGA, GbE NIC,Tower Case | 12.000.000 |
| Stabilizer | 30.000.000 |
| UPS | 40.000.000 |
| **Total** | **82.000.000** |

Tabel 6 Biaya pengadaan peralatan jaringan

| Alat Jaringan | Harga |
|---|---|
| Switch 3Com 3C16470 SuperStack III Baseline 10/100 Mbps 16 port MDI/MDIX 10BASE-T/100 BASE-TX 2unit @ Rp. 1.400.000 | 2.800.000 |
| Kabel UTP Kategori 6 | 1.050.000 |
| RJ-45 + connector shield | 750.000 |
| Network Interface Card | 1.200.000 |
| **Total** | **5.800.000** |

Tabel 7 Biaya pendukung

| Biaya pendukung | Harga |
|---|---|
| Mengadakan Seminar | 5.000.000 |
| Peralatan Alat Tulis Kantor | 2.000.000 |
| Lain-lain | 5.000.000 |
| Lampu cadangan | 500.000 |
| Rak Komputer Server | 1.900.000 |
| **Total** | **14.400.000** |

Total keseluruhan biaya untuk pengembangan proyek ini terlihat pada tabel 8 dan akan menjadi nilai investasi proyek ini sewaktu melakukan penghitungan analisa manfaat/biaya dengan metode simple ROI (Return On Investment).

Tabel 8 Total keseluruhan biaya

| Biaya-biaya | Harga |
|---|---|
| Upah staf proyek | 136.500.000 |
| Hardware | 82.000.000 |
| Alat Jaringan | 5.800.000 |
| Biaya pendukung | 14.400.000 |
| **Total Biaya** | **238.700.000** |

## 4 Analisa biaya

Sebelum melakukan perhitungan dengan menggunakan metode simple ROI (Return On Investment) ini perlu bagi kita untuk menganalisa biaya berjalan yang mendukung proyek ini serta biaya operasional yang selama ini dapat dihemat.

1) Biaya Berjalan

Tabel 9 memperlihatkan biaya berjalan proyek dimana biaya ini adalah biaya yang dibutuhkan selama 5 tahun kedepan untuk mendukung kelancaran pelaksanaan proyek ini. Dimana tahun ke-1 tidak ada sama sekali biaya dan dimulai tahun ke-2 dan biaya berkurang 10% pada tahun berikutnya, hal ini didasarkan pada asumsi inflasi Rupiah dan penyesuaian UMR (Upah Minimum Regional) sebesar 10% pertahun.

Tabel 9 Biaya berjalan proyek

| | Biaya | Tahun ke-1 | Tahun ke-2 | Tahun ke-3 | Tahun ke-4 | Tahun ke-5 |
|---|---|---|---|---|---|---|
| A | Pemeliharaan Aplikasi | | | | | |
| | 1. Penyempurnaan Sistem | 0 | 30,150,000 | 27,135,000 | 24,421,500 | 21,979,350 |
| B | Pemeliharaan Perangkat Keras | | | | | |
| | 1. Peningkatan memory server | 0 | 0 | 16,000,000 | 14,400,000 | 12,960,000 |
| | 2. Peningkatan hardisk server | 0 | 0 | 24,000,000 | 21,600,000 | 19,440,000 |
| | **Total Biaya berjalan** | **0** | **30,150,000** | **67,135,000** | **60,421,500** | **54,379,350** |

2) Biaya operasional pembuatan laporan

Biaya operasional pembuatan laporan ini meliputi :

a. Administrasi pembuatan laporan

Biaya administrasi pembuatan laporan ini didapatkan dari bagian keuangan, dimana masing-masing biaya merupakan biaya per tahun yang harus dikeluarkan oleh bagian keuangan untuk mendukung pembuatan laporan-laporan yang dibutuhkan oleh manajemen tingkat atas perguruan tinggi. Biaya administrasi pembuatan laporan meliputi biaya pengadaan :

a) Kertas

Dengan harga per rim sekitar Rp. 30.000, didapatkan bahwa dalam 1 tahun menghabiskan sekitar 12 rim kertas. Kertas ini dipakai untuk mencetak hasil-hasil laporan dan memperbanyak hasil laporan untuk dijadikan bahan rapat.

b) Tinta printer

Dalam 1 tahun dibutuhkan 1 toner printer Hp Q2613A untuk printer Hp Laser Jet 1300 yang berharga Rp. 1.100.000.

c) Tinta foto copy

Dalam 1 tahun dibutuhkan setengah toner mesin foto copy Canon tipe NPG-8 untuk mesin foto copy merk Canon tipe 6130. Toner mesin foto copy ini berharga 900.000.

d) Honor panitia

Sesuai dengan data yang didapatkan dari bagian keuangan pelaksanaan pembuatan laporan untuk manajemen tingkat atas ini membutuhkan pembentukan panitia yang mana harus mengeluarkan honor untuk panitia yang mencapai 12.000.000 per tahun.

e) Dana rapat

Sesuai dengan data yang didapatkan dari bagian keuangan dibutuhkan dana mencapai 35.000.000 untuk panitia dalam mengadakan rapat-rapat, yang kadangkala harus mengadakan rapat ke luar kota.

b. Tenaga pembuatan laporan

Pembuatan laporan biasanya melibatkan manajemen tingkat atas seperti Dekan, Ketua Program Studi, sekretaris program studi dan dibantu oleh beberapa staf pengolahan data seperti seorang sistem analis dan 2 orang programer . Selain itu didukung oleh 2 orang staf biasa yang bertugas membantu dalam penyiapan laporan dan memperbanyak laporan. Berikut ini adalah gaji per bulan untuk masing-masing staf selain manajemen tingkat atas yaitu Sistem analis 1 orang dengan gaji 5 juta per bulan, Programer 2 orang dengan gaji masing-masing 3 juta per bulan dan Staf 2 orang dengan gaji masing-masing 2 juta per bulan.

Tabel 10 memperlihatkan berapa biaya operasional yang dibutuhkan pada sistem yang berjalan sebelum diterapkannya proyek ini dan diharapkan biaya ini dapat dikurangi. Biaya-biaya ini diasumsikan mengalami kenaikan 10% per tahunnya, hal ini didasarkan pada asumsi inflasi Rupiah dan penyesuaian UMR (Upah Minimum Regional) sebesar 10% pertahun.

Tabel 10 Biaya operasional pembuatan laporan

| Jenis Biaya | Tahun ke-1 | Tahun Ke-2 | Tahun ke-3 | Tahun ke-4 | Tahun ke-5 |
|---|---|---|---|---|---|
| Biaya administrasi Pembuatan Laporan | | | | | |
| Kertas | 360,000 | 396,000 | 435,600 | 479,160 | 527,076 |
| Tinta Printer | 1,100,000 | 1,210,000 | 1,331,000 | 1,464,100 | 1,610,510 |
| Tinta FotoCopy | 450,000 | 495,000 | 544,500 | 598,950 | 658,845 |
| Honor panitia | 12,000,000 | 13,200,000 | 14,520,000 | 15,972,000 | 17,569,200 |
| Rapat | 35,000,000 | 38,500,000 | 42,350,000 | 46,585,000 | 51,243,500 |
| Biaya tenaga Pembuatan Laporan | | | | | |
| Sistem Analis (1 orang) | 60,000,000 | 66,000,000 | 72,600,000 | 79,860,000 | 87,846,000 |
| Programer (2 orang) | 72,000,000 | 79,200,000 | 87,120,000 | 95,832,000 | 105,415,200 |
| Staf (2 orang) | 48,000,000 | 52,800,000 | 58,080,000 | 63,888,000 | 70,276,800 |
| **Total Biaya Operasional** | **228,910,000** | **251,801,000** | **276,981,100** | **304,679,210** | **335,147,131** |

## 5 Manfaat penghematan biaya operasional

Penghematan biaya operasional pembuatan laporan ini meliputi :

1) Administrasi pembuatan laporan

Penghematan administrasi pembuatan laporan meliput :

a. Kertas

Dengan adanya Data Warehouse pencetakan laporan yang belum sempurna tidak diperlukan, manajemen dapat mengakses laporan secara langsung melalui aplikasi dan dapat menghemat 75% penggunaan kertas.

b. Tinta printer

Dengan adanya pengurangan pencetakan laporan yang belum sempurna dan dapat menghemat 75% penggunaan kertas, maka otomatis menghemat penggunaan tinta printer sampai 75%.

c. Tinta foto copy

Dengan adanya pengurangan pencetakan laporan yang belum sempurna dan adanya fasilitas bagi manajemen tingkat atas untuk mengakses laporan secara langsung melalui aplikasi maka memperbanyak laporan yang belum sempurna untuk kepentingan rapat dapat dikurangi sampai 75%.

d. Honor panitia

Dengan adanya adanya fasilitas bagi manajemen tingkat atas untuk mengakses laporan secara langsung melalui aplikasi, maka tidak diperlukan pembentukan panitia pada setiap pembuatan laporan untuk manajemen tingkat atas. Sehingga pengeluaran biaya honor panitia dapat dikurangi sampai 100%.

e. Dana rapat

Dengan adanya adanya fasilitas bagi manajemen tingkat atas untuk mengakses laporan secara langsung melalui aplikasi dan tidak diperlukannya pembentukan panitia maka biaya rapat yang kadangkala harus mengadakan rapat ke luar kota dapat dikurangi sampai 75%.

2) Tenaga pembuatan laporan

Dengan adanya adanya fasilitas bagi manajemen tingkat atas untuk mengakses laporan secara langsung melalui aplikasi maka ada kemungkinan bagi BSI (Biro Sistem Informasi) untuk mengurangi jumlah staff seperti : Sistem analis 1 orang dengan gaji 5 juta per bulan, Programer 2 orang dengan gaji masing-masing 3 juta per bulan dan staff sebanyak 2 orang dengan gaji masing-masing 2 juta per bulan

Tabel 11 memperlihatkan penghematan dari biaya operasional diatas sesuai dengan prosentase penghematan masing-masing.

Tabel 11 Penghematan biaya operasional pembuatan laporan

| Jenis Penghematan | Tahun ke-1 | Tahun ke-2 | Tahun Ke-3 | Tahun ke-4 | Tahun ke-5 |
|---|---|---|---|---|---|
| Biaya Administrasi Pembuatan Laporan | | | | | |
| Kertas (75%) | 270,000 | 297,000 | 326,700 | 359,370 | 395,307 |
| Tinta Printer (75%) | 825,000 | 907,500 | 998,250 | 1,098,075 | 1,207,883 |
| Tinta FotoCopy (75%) | 337,500 | 371,250 | 408,375 | 449,213 | 494,134 |
| Honor panitia (100%) | 12,000,000 | 13,200,000 | 14,520,000 | 15,972,000 | 17,569,200 |
| Biaya rapat (75%) | 26,250,000 | 28,875,000 | 31,762,500 | 34,938,750 | 38,432,625 |
| Tenaga Pembuatan Laporan | | | | | |
| Sistem Analis (1 orang) | 60,000,000 | 66,000,000 | 72,600,000 | 79,860,000 | 87,846,000 |
| Programer (2 orang) | 72,000,000 | 79,200,000 | 87,120,000 | 95,832,000 | 105,415,200 |
| Staf (2 orang) | 48,000,000 | 52,800,000 | 58,080,000 | 63,888,000 | 70,276,800 |
| **Total Penghematan** | **219,682,500** | **241,650,750** | **265,815,825** | **292,397,408** | **321,637,148** |

## 6 Manfaat meningkatkan jumlah mahasiswa baru

Melalui Data Warehouse diharapkan akan membantu perguruan tinggi untuk lebih matang lagi dalam membuat keputusan guna meningkatkan jumlah mahasiswa baru yang otomatis akan meningkatkan pendapatan [12][1]. Sesuai dengan data penerimaan mahasiswa baru tabel 12 memperlihatkan contoh jumlah penerimaan mahasiswa baru mencapai puncaknya pada tahun 1999 dengan jumlah 2390 mahasiswa baru. Penerimaan mahasiswa baru dari tahun 1999 hingga tahun 2005 mengalami penurunan antara 8% sampai 46% dari tahun sebelumnya, walaupun pada tahun 2004 sempat mengalami kenaikan 10% dari tahun sebelumnya.

Tabel 12 Data penerimaan mahasiswa baru

| Tahun | Program Studi | | | | Total | Selisih | Prosentase (%) Selisih |
|---|---|---|---|---|---|---|---|
| | TI | SI | SK | AK | | | |
| 1986 | 520 | 892 | 37 | 113 | 1,562 | | |
| 1987 | 169 | 599 | 60 | 270 | 1,098 | -464 | -30% |
| 1988 | 156 | 389 | 86 | 260 | 891 | -207 | -19% |
| 1989 | 79 | 580 | 59 | 124 | 842 | -49 | -5% |
| 1990 | 142 | 1,072 | 51 | 126 | 1,391 | 549 | 65% |
| 1991 | 202 | 1,078 | 33 | 71 | 1,384 | -7 | -1% |
| 1992 | 261 | 1,038 | 71 | 126 | 1,496 | 112 | 8% |
| 1993 | 205 | 835 | 148 | 339 | 1,527 | 31 | 2% |
| 1994 | 254 | 898 | 95 | 318 | 1,565 | 38 | 2% |
| 1995 | 402 | 1,185 | 122 | 328 | 2,037 | 472 | 30% |
| 1996 | 349 | 1,031 | 111 | 339 | 1,830 | -207 | -10% |
| 1997 | 525 | 1,205 | 93 | 304 | 2,127 | 297 | 16% |
| 1998 | 501 | 1,081 | 156 | 256 | 1,994 | -133 | -6% |
| 1999 | 645 | 1,268 | 152 | 325 | 2,390 | 396 | 20% |
| 2000 | 697 | 1,108 | 117 | 274 | 2,196 | -194 | -8% |
| 2001 | 673 | 861 | 46 | 197 | 1,777 | -419 | -19% |
| 2002 | 559 | 733 | 35 | 237 | 1,564 | -213 | -12% |

| | | Program Studi | | | Total | Selisih | Prosentase (%) Selisih |
|---|---|---|---|---|---|---|---|
| **Tahun** | **TI** | **SI** | **SK** | **AK** | | | |
| 2003 | 455 | 534 | 32 | 193 | 1,214 | -350 | -22% |
| 2004 | 580 | 543 | 43 | 166 | 1,332 | 118 | 10% |
| 2005 | 312 | 295 | 33 | 79 | 719 | -613 | -46% |
| **Total** | 7,686 | 17,225 | 1,580 | 4,445 | 30,936 | | |

Diharapkan dengan dibuatkannya Data Warehouse manajemen tingkat atas dapat meningkatkan jumlah penerimaan mahasiswa baru sampai 20% per tahun. Angka 20% ini merupakan nilai rata-rata dari tingkat kenaikan jumlah penerimaan mahasiswa baru (65%+8%+2%+2%+30%+16%+20%+10%)/8 = 19.125 %.

Berikut ini tabel 13 merupakan biaya yang harus dikeluarkan oleh mahasiswa baru dan tabel 14 sebagai tabel biaya sumbangan gedung.

Tabel 13 Biaya mahasiswa baru

| Biaya untuk mahasiswa baru | Harga |
|---|---|
| Daftar Ulang | 200.000 |
| SKS (per sks 65000) 20 sks | 1.300.000 |
| Operasional pendidikan | 1.960.000 |
| Dana Kemahasiswaan | 15.000 |
| Koperasi Mahasiswa | 10.000 |
| Paket Mahasiswa baru | 500.000 |
| **Total** | **3.985.000** |

Tabel 14 Biaya sumbangan gedung

| Sumbangan Gedung | Harga |
|---|---|
| Grade A | 6.000.000 |
| Grade B | 6.500.000 |
| Grade C | 7.000.000 |
| Grade D | 8.000.000 |
| **Rata-rata Sumbangan Gedung** | **6.875.000** |

Berdasarkan biaya yang harus dikeluarkan mahasiswa diatas didapatlah rata-rata biaya semester awal : Rp. 3,985,000 + Rp. 6,875,000 = Rp. 10,860,000. Biaya rata-rata semester awal ini dikurangkan dengan Dana kemahasiswaan, koperasi mahasiswa dan paket mahasiswa baru yang memang masing-masing telah dialokasikan menurut kebutuhannya. Jadi Rp. 10,860,000 – Rp. 525.000 = Rp. 10,335,000. Berdasarkan data rata-rata dari bagian keuangan bahwa penggunaan dari rata-rata biaya yang harus dikeluarkan mahasiswa yang dipakai untuk operasional seperti gaji dosen dan karyawan, dan biaya lainnya mencapai 40%.

Sehingga Manfaat bersih yang didapat dari setiap mahasiswa Rp. 10,335,000 - Rp. 4,134,000 = Rp. 6,201,000 dan mengalami kenaikan dengan asumsi 5% per tahunnya agar harga kenaikan biaya kuliah selalu melihat kepada persaingan harga pendidikan sekolah tinggi swasta lainnya. Tabel 15 akan menjelaskan perkiraan pendapatan 5 tahun ke depan dari peningkatan jumlah mahasiswa baru.

Tabel 15 Perkiraan pendapatan dari peningkatan jumlah mahasiswa baru

| **Program Studi** | **Tahun ke-1** | | **Tahun ke-2** | | **Tahun ke-3** | | **Tahun ke-4** | | **Tahun ke-5** | |
|---|---|---|---|---|---|---|---|---|---|---|
| **Rata-rata biaya smst 1** | **2.067.000** | | **2.170.350** | | **2.278.868** | | **2.392.811** | | **2.512.451** | |
| | **20%** | **Rp** | **20%** | **Rp** | **20%** | **Rp** | **20%** | **Rp** | **20%** | **Rp** |
| **TI** | 62 | 128.154.000 | 74 | 328.074.240 | 89 | 579.973.742 | 107 | 814.067.015 | 129 | 1.109.024.539 |
| **SI** | 59 | 121.953.000 | 71 | 312.199.680 | 85 | 551.910.497 | 102 | 774.676.676 | 122 | 1.055.362.062 |
| **SK** | 7 | 14.469.000 | 8 | 37.040.640 | 10 | 65.480.906 | 12 | 91.910.792 | 15 | 125.212.448 |
| **AK** | 16 | 33.072.000 | 19 | 84.664.320 | 23 | 149.670.643 | 28 | 210.081.810 | 33 | 286.199.881 |
| **Total** | 144 | 297.648.000 | 173 | 761.978.880 | 207 | 1.347.035.789 | 249 | 1.890.736.294 | 299 | 2.575.798.930 |

Pada tabel 15 jumlah mahasiswa baru mengalami kenaikan 20% per program studinya pada tahun sesudah tahun ke-1 berdasarkan jumlah kenaikan jumlah mahasiswa baru 1 tahun sebelumnya. Nilai manfaat bersih per program studi yang didapat untuk tahun pertama hanya diambil dari jumlah biaya rata-rata semester 1 dikalikan dengan jumlah mahasiswa baru. Sedangkan nilai manfaat bersih mulai tahun ke-2 sampai tahun ke-5 dijumlahkan dengan jumlah rupiah yang didapat tahun sebelumnya dari mahasiswa yang mendaftar pada tahun sebelumnya dan biaya yang harus dikeluarkan untuk mengambil kuliah pada 2 semester berikutnya.

## 7 Peningkatan produktivitas manajemen tingkat atas

Melalui Data Warehouse diharapkan akan membantu perguruan tinggi untuk meningkatkan produktivitas manajemen tingkat atas. Diharapkan dengan dibuatkannya sebuah aplikasi yang dapat diakses langsung oleh manajemen tingkat atas maka komposisi kerja manajemen tingkat atas dapat lebih efektif sehingga produktivitas kerja manajemen tingkat atas akan lebih meningkat. Agak sulit untuk mengukur prosentase kerja masing-masing jabatan dan berdasarkan hasil wawancara dengan pihak manajemen tingkat atas maka tabel 16 adalah tabel yang menggambarkan komposisi kerja manajemen tingkat atas saat ini yaitu dimana Dekan = 50%, Ketua Program Studi = 60% dan Sekretaris Program Studi = 70%.

Tabel 16 Produktivitas manajemen tingkat atas sebelum implementasi proyek Data Warehouse

| **Jabatan** | **Jmlh Orang** | **Dekan** | | **Ketua Program Studi** | | **Sekretaris Program Studi** | | **Non** | | **Total** | |
|---|---|---|---|---|---|---|---|---|---|---|---|
| | | | **96.000.000** | | **72.000.000** | | **48.000.000** | | **0** | | **216.000.000** |
| | | **%** | **Rp** | **%** | **Rp** | **%** | **Rp** | **%** | **Rp** | **%** | **Rp** |
| Dekan | 1 | 50 | 48.000.000 | 30 | 21.600.000 | 15 | 7.200.000 | 5 | 0 | 100 | 76.800.000 |
| Ketua Program studi | 4 | 10 | 38.400.000 | 60 | 172.800.000 | 15 | 28.800.000 | 15 | 0 | 100 | 40.000.000 |
| Sekretaris Program Studi | 1 | 5 | 4.800.000 | 10 | 7.200.000 | 70 | 33.600.000 | 15 | 0 | 100 | 45.600.000 |
| Total Produktivitas | | | | | | | | | | | 362.400.000 |
| Biaya Total yang dikeluarkan | | | | | | | | | | | 432.000.000 |
| Kerugian Waktu produktif | | | | | | | | | | | 69.600.000 |

Tabel 17 Produktivitas manajemen tingkat atas setelah implementasi proyek Data Warehouse

| **Jabatan** | **Jumlah Orang** | **Dekan** | | **Ketua Program Studi** | | **Sekretaris Program Studi** | | **Non** | | **Total** | |
|---|---|---|---|---|---|---|---|---|---|---|---|
| | | | **96.000.000** | | **72.000.000** | | **48.000.000** | | **0** | | **216.000.000** |
| | | **%** | **Rp** | **%** | **Rp** | **%** | **Rp** | **%** | **Rp** | **%** | **Rp** |
| Dekan | 1 | 65 | 62.400.000 | 25 | 18.000.000 | 10 | 4.800.000 | 0 | 0 | 100 | 85.200.000 |
| Ketua Program Studi | 4 | 15 | 57.600.000 | 75 | 216.000.000 | 5 | 9.600.000 | 5 | 0 | 100 | 283.200.000 |
| Sekretaris Program Studi | 1 | 2 | 1.920.000 | 8 | 5.760.000 | 85 | 40.800.000 | 5 | 0 | 100 | 48.480.000 |
| Total Produktivitas | | | | | | | | | | | 416.880.000 |
| Biaya Total yang dikeluarkan | | | | | | | | | | | 432.000.000 |
| Kerugian Waktu produktif | | | | | | | | | | | 15.120.000 |

Tabel 17 adalah tabel yang mengkomposisikan peningkatan produktivitas kerja manajemen tingkat atas setelah pengimplementasian proyek Data Warehouse, dimana Dekan dari 50% menjadi 65 %, Ketua Program Studi dari 60% menjadi 75% dan Sekretaris Program Studi dari 70% menjadi 85 %. Peningkatan prosentase masing-masing jabatan tersebut didapat berdasarkan hasil wawancara dengan pihak manajemen tingkat atas, yang pada akhirnya diharapkan produktfitas manajemen tingkat atas dapat tercapai sesuai dengan penugasan jabatan manajemen tingkat atas tersebut masing-masing dan diharapkan keputusan-keputusan strategis yang akan ditentukan oleh manajemen tingkat atas ini akan lebih akurat dan terarah pada sasarannya.

Dari hasil perbandingan kerugian waktu produktif kedua tabel maka didapatlah rekapitulasi efisiensi produktivitas kerja tahun pertama :

Rp. 69.600.000 – Rp. 15.120.000 = Rp. 54.480.000.

Setiap tahun seterusnya lima tahun kedepan akan mengalami peningkatan 10% dengan asumsi adanya kenaikan gaji 10% per tahun, hal ini didasarkan pada asumsi inflasi Rupiah dan penyesuaian UMR (Upah Minimum Regional) sebesar 10% pertahun yang terlihat pada tabel 18.

Tabel 18 Rekapitulasi efisiensi produktivitas kerja

| Tahun ke-1 | Tahun ke-2 | Tahun ke-3 | Tahun ke-4 | Tahun ke-5 |
|---|---|---|---|---|
| 54,480,000 | 59,928,000 | 65,920,800 | 72,512,880 | 79,764,168 |

## 8 Perhitungan simple ROI

Perhitungan ROI (Return On Investment) ini digunakan untuk melihat apakah proyek Data Warehouse pada perguruan tinggi ini layak untuk diimplementasikan. Perhitungan untuk mendapatkan nilai arus kas bersih yaitu dengan menjumlah semua manfaat yang diperoleh dan dikurangkan dengan seluruh biaya yang terlihat pada table 19. Nilai arus kas bersih ini yang dipakai untuk menghitung nilai prosentase ROI (Return On Investment). Berikut ini penjelasan sesuai dengan perhitungan diatas yaitu :

Tabel 19 Arus kas bersih

| Manfaat dan Biaya | Harga |
|---|---|
| Penerimaan Mahasiswa baru | 20,619,593,679 |
| Peningkatan produktivitas manajemen tingkat atas | 332,605,848 |
| | -----------------+ |
| Manfaat Ekonomi Bersih | 20,952,199,527 |
| Pengurangan Biaya Operasional | 1,341,183,631 |
| | -----------------+ |
| Pendapatan Sebelum Pajak | 22,293,383,158 |
| Biaya Berjalan | 212,085,850 |
| | ---------------- - |
| **Arus Kas Bersih** | **22,081,297,308** |

Manfaat penerimaan Mahasiswa baru didapat dari penjumlahan pada tabel 15.

```
    892,944,000
  2,285,936,640
  4,041,107,366
  5,672,208,882
  7,727,396,791  +
 20,619,593,679
```

Manfaat peningkatan produktivitas manajemen tingkat atas didapat dari penjumlahan pada tabel 18 yaitu :

```
 54,480,000
 59,928,000
 65,920,800
 72,512,880
 79,764,168  +
332,605,848
```

Sedangkan manfaat pengurangan biaya operasional didapat dari penjumlahan pada tabel 11 yaitu

```
  219,682,500
  241,650,750
  265,815,825
  292,397,408
  321,637,148  +
1,341,183,631
```

Sedangkan biaya berjalan didapat dari penjumlahan total biaya berjalan proyek pada tabel 9.

0
30,150,000
67,135,000
60,421,500
54,379,350 +
212,085,850

Pada akhirnya nilai prosentase ROI (Return On Investment) akan didapat dari Total arus kas bersih dibagi dengan 5 tahun dan dibagi lagi dengan nilai investasi bersih proyek [8][11]. Berikut ini merupakan rumus untuk mencarikan prosentase ROI.

Prosentase ROI:

Arus kas bersih
5 tahun
Nilai Investasi proyek

Dimana Arus kas bersih Rp. **22,081,297,308** dan Nilai investasi proyek Rp. **238,700,000**.

Nilai Arus kas bersih sebesar 22,081,297,308 didapat dari tabel 19 dan Nilai investasi proyek sebesar 238,700,000 didapat dari tabel 8.

22,081,297,308 / 5 / 238,700,000 = 1850,13 %

## 9 Kesimpulan

Berdasarkan data ROI (Return On Investment) yang mencapai 1850.13 % dan total arus kas bersih yang mencapai Rp. 22,081,297,308 maka sudah dipastikan bahwa proyek Data Warehouse pada perguruan tinggi layak untuk diimplementasikan.

Nilai prosentase ROI akan dapat berubah sesuai dengan nilai perhitungan investasi, biaya-biaya yang dikeluarkan dan kejelian untuk mengkuantifikasi hal-hal yang kualitatif sehingga meningkatkan manfaat-manfaat yang sifatnya tidak dapat diukur menjadi dapat diukur.

Proyek teknologi informasi yang cenderung *cost centre* akan dapat dicermati dengan mengelola sumber-sumber manfaat yang ada sehingga dapat dikuantifikasi dan diukur, sehingga meningkatkan keinginan manajemen tingkat atas untuk melakukan investasi teknologi informasi.

Proyek data warehouse yang cenderung mempunyai nilai investasi tinggi dan hanya digunakan oleh manajemen tingkat atas membuat keengganan pembuat keputusan untuk mengimplementasikan proyek data warehouse. Namun diharapkan penggunaan metode simple ROI ini dapat digunakan sebagai salah satu alat pendukung bagi pengambil keputusan untuk mengimplementasikan proyek data warehouse.

Metode analisa manfaat/biaya yang lain dapat digunakan untuk mempertegas hasil simpel ROI yang didapat ataupun kebalikannya dapat memperkuat metode analisa manfaat/biaya yang lainnya.

Analisa dampak investasi teknologi informasi ini diterapkan pada perguruan tinggi swasta, dikarenakan perguruan tinggi swasta harus bersaing untuk mendapatkan calon mahasiswa baru sebagai salah satu sumber pendapatan bagi perguruan tinggi swasta. Perguruan tinggi negeri biasanya tidak perlu bersaing untuk mendapatkan calon mahasiswa baru, dikarenakan jumlah kursi yang disediakan lebih sedikit dibandingkan dengan peminatnya. Namun tidak tertutup kemungkinan jika diterapkan pada perguruan tinggi negeri dengan perubahan-perubahan pada perhitungan investasi yang disesuaikan dengan kebutuhan perguruan tinggi negeri.

## 10 Daftar Pustaka